\documentclass[conference]{IEEEtran}
\IEEEoverridecommandlockouts

\usepackage{cite}
\usepackage{amsmath,amssymb,amsfonts}
\usepackage{algorithmic}
\usepackage{graphicx}
\usepackage{textcomp}
\usepackage{xcolor}
\usepackage{booktabs}
\usepackage{wrapfig}

\usepackage{caption}

\def\BibTeX{{\rm B\kern-.05em{\sc i\kern-.025em b}\kern-.08em
    T\kern-.1667em\lower.7ex\hbox{E}\kern-.125emX}}
\begin{document}

\title{pMSz: A Distributed Parallel Algorithm for Correcting Extrema and Morse-Smale Segmentations in Lossy Compression}

\author{
\IEEEauthorblockN{Yuxiao Li}
\IEEEauthorblockA{\textit{The Ohio State University}\\
Columbus, OH, USA\\
li.14025@osu.edu}\\
\IEEEauthorblockN{Sheng Di}
\IEEEauthorblockA{\textit{Argonne National Laboratory}\\
Lemont, IL, USA\\
sdi1@anl.gov}
\and
\IEEEauthorblockN{Mingze Xia}
\IEEEauthorblockA{\textit{Oregon State University}\\
Corvallis, OR, USA\\
xiaml@oregonstate.edu}\\
\IEEEauthorblockN{Hemant Sharma}
\IEEEauthorblockA{\textit{Argonne National Laboratory}\\
Lemont, IL, USA\\
hsharma@anl.gov}
\and
\IEEEauthorblockN{Xin Liang}
\IEEEauthorblockA{\textit{Oregon State University}\\
Corvallis, OR, USA\\
linuxin@oregonstate.edu}\\
\IEEEauthorblockN{Dishant Beniwal}
\IEEEauthorblockA{\textit{Argonne National Laboratory}\\
Lemont, IL, USA\\
dbeniwal@anl.gov}
\and
\IEEEauthorblockN{Bei Wang}
\IEEEauthorblockA{\textit{University of Utah}\\
Salt Lake City, UT, USA\\
beiwang@sci.utah.edu}\\
\IEEEauthorblockN{Franck Cappello}
\IEEEauthorblockA{\textit{Argonne National Laboratory}\\
Lemont, IL, USA\\
cappello@mcs.anl.gov}
\and
\IEEEauthorblockN{Robert Underwood}
\IEEEauthorblockA{\textit{Argonne National Laboratory}\\
Lemont, IL, USA\\
runderwood@anl.gov}
\and
\IEEEauthorblockN{Hanqi Guo}
\IEEEauthorblockA{\textit{The Ohio State University}\\
Columbus, OH, USA\\
guo.2154@osu.edu}
}
\maketitle

\begin{abstract}
Lossy compression, widely used by scientists to reduce data from simulations, experiments, and observations, can distort features of interest even under bounded error. Such distortions may compromise downstream analyses and lead to incorrect scientific conclusions in applications such as combustion and cosmology. This paper presents a distributed and parallel algorithm for correcting topological features, specifically, piecewise linear Morse–Smale segmentations (PLMSS), which decompose the domain into monotone regions labeled by their corresponding local minima and maxima. While a single-GPU algorithm (MSz) exists for PLMSS correction after compression, no methodology has been developed that scales beyond a single GPU for extreme-scale data. We identify the key bottleneck in scaling PLMSS correction as the parallel computation of integral paths, a communication-intensive computation that is notoriously difficult to scale. Instead of explicitly computing and correcting integral paths, our algorithm simplifies MSz by preserving steepest ascending and descending directions across all locations, thereby minimizing interprocess communication while introducing negligible additional storage overhead. With this simplified algorithm and relaxed synchronization, our method achieves over 90\% parallel efficiency on 128 GPUs on the Perlmutter supercomputer for real-world datasets.
\end{abstract}


\section{Introduction}
The rapid advancement of high-performance computing (HPC) and scientific instruments has led to explosive growth in scientific data volumes across domains such as cosmology, combustion, and climate modeling, posing significant challenges for storage, I/O, and analysis pipelines~\cite{Di2024}. For example, a trillion-particle cosmology simulation using the HACC framework can generate over 20 PB of raw data in a single run~\cite{Kettimuthua17}. 
To this end, compression techniques, especially lossy compression that significantly reduces the size of scientific data, have been widely used in scientific domains such as climate modeling and cosmology simulations~\cite{Tao_2017}, where large-scale datasets are generated routinely and require efficient storage and transfer.
In contrast to general-purpose compression, scientific applications often require preserving data fidelity within strict error bounds to ensure the validity of downstream analyses, for which error-bounded lossy compression techniques have been widely adopted to achieve high compression ratios with strictly controlled errors in the decompressed data~\cite{sz, sz3, Tao_2017, Kai2021, Kai2020, zfp, FPZIP}. To meet the growing data demands of large-scale simulations, many of these compressors have been further extended with parallel implementations to efficiently handle ever-increasing data scales~\cite{sz3, zfp, huang2023cuszp, huang2024cuszp2}.

\begin{figure}[htb]
    \centering
    \includegraphics[width=\linewidth]{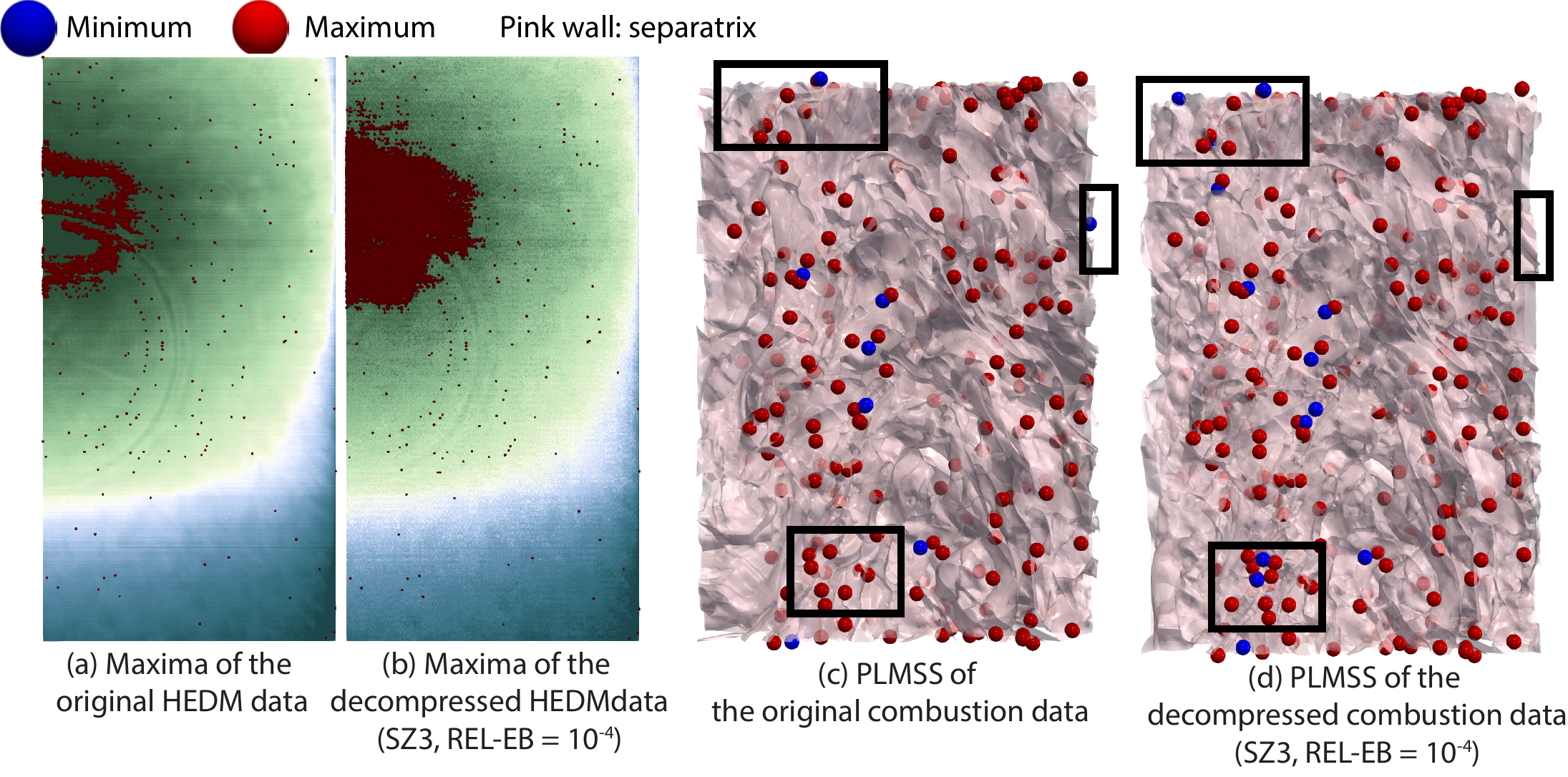}
    \caption{Impacts of lossy compression (SZ3) on topological features. (a) and (b): Positions of local maxima in the original and decompressed High-Energy Diffraction Microscopy (HEDM) datasets (SZ3, relative error bound $10^{-4}$). Noticeable distortion of maxima can be observed in the decompressed data, which may lead to inaccuracies in subsequent peak analysis. (c) and (d): Piecewise linear Morse–Smale segmentations (PLMSS) of the original and decompressed combustion datasets. Distorted segmentations, as highlighted in black boxes, may influence downstream analysis of flow and combustion structures.}
    \label{fig:distortion}
\end{figure}
\begin{figure}
    \centering
    \includegraphics[width=\linewidth]{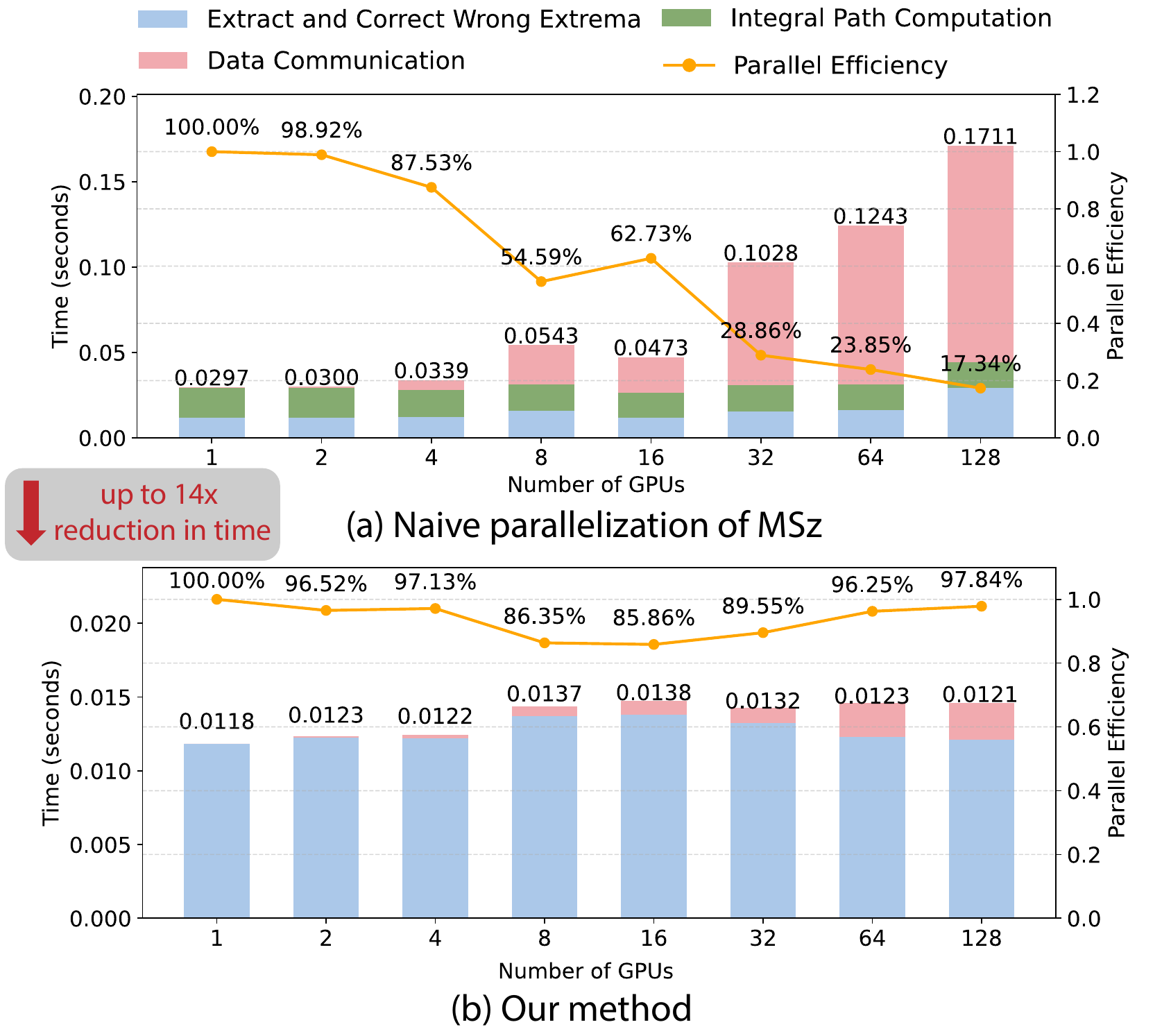}
    \caption{Comparison of parallel efficiency between the direct distributed parallelization of MSz and our proposed method under weak scaling on the Perlin noise dataset (each GPU processes a fixed $512^3$ problem size). (a) Direct parallelization of MSz suffers from poor scalability due to the global dependencies introduced by integral path computation in each iteration, resulting in less than 20\% efficiency at 16 GPUs. (b) Our method eliminates global integral path tracing through a relaxed sufficient condition and local operations, achieving over 90\% parallel efficiency at 128 GPUs.}
    \label{fig:msz}
\end{figure}
While many error-bounded lossy compressors have been extended with parallel implementations to efficiently handle ever-growing data scales~\cite{sz3, zfp, huang2023cuszp, huang2024cuszp2}, there is a lack of distributed-parallel algorithms to preserve scientific features or topological structures, which are often essential for ensuring the correctness of downstream analyses.
In applications such as chemistry~\cite{Bhatia_2018, Günther_2014}, cosmology~\cite{Shivashankar_2015}, and combustion~\cite{Dora_2013, Chen2009}, the correctness of analysis depends not only on numerical accuracy but also on topological features (e.g., extrema and Morse–Smale segmentations~\cite{MSS}) that offer abstraction and insights into the complex structures of data. Inconsistencies in topological features (as shown in Figure~\ref{fig:distortion}) between the original and decompressed data may influence downstream analyses and lead to erroneous scientific results. For example, in the analysis of high-energy diffraction microscopy (HEDM) data, distorted peaks, corresponding to grain orientations and strain states in crystalline materials, may lead to incorrect grain identification and misinterpretation of internal strain distributions, as shown in Figure~\ref{fig:distortion} (a) and (b); in combustion analysis, distorted segmentations may misrepresent flame boundaries (as shown in Figure~\ref{fig:distortion} (c) and (d)), leading to flawed interpretations of burning zones and inaccurate assessments of the underlying physical processes~\cite{Chen2009}.

Although several feature-preserving approaches have been proposed~\cite{li_msz,yan2023toposz,gorski2025general,Xia24,Xia25,Liang_2020}, existing methods provide only limited scalability due to their reliance on shared-memory parallelism designs. 
For example, MSz~\cite{li_msz} supports single GPU-based parallelism for preserving piecewise linear Morse–Smale segmentations (PLMSS), but it fails to scale efficiently on distributed-memory systems because it relies on integral path computation, i.e.,~trajectories obtained by following the steepest ascent or descent direction from each data point to its corresponding local extremum. In our experiment, with a naive parallelization of MSz over distributed processes, as shown in Figure~\ref{fig:msz}(a), the parallel efficiency of the naive parallelization of MSz drops below 20\% at 128 GPUs under weak scaling. The drop in parallel efficiency in MSz is primarily attributed to parallel integral line computation, a notoriously difficult problem to scale efficiently~\cite{MSS}. Specifically, in each iteration, MSz recomputes the PLMSS by performing integral path tracing for every data point, following the gradient direction either ascending or descending until convergence to a local maximum or minimum; the integral path computation itself requires frequent communication and synchronization that are difficult to parallelize efficiently~\cite{MSS} and accounts for up to 60\% of the total computation time (the green bar in Figure~\ref{fig:msz}(a)). The computation of integral paths in each iteration also causes the paths that traverse across block boundaries to require boundary data exchanges between neighboring processes to ensure correctness, leading to substantial communication overhead, as reflected by the increasing communication time in Figure~\ref{fig:msz}(a).

In this paper, we significantly improve the parallel execution performance and scalability of MSz, especially in the distributed memory environment. To address the scalability bottlenecks in the integral path tracing, we relax the requirements for PLMSS preservation to avoid global integral line tracing and reduce data communication frequency. Specifically, the dataset is partitioned into blocks and processed in a distributed-memory environment, with each processor responsible for its assigned block. In each iteration, the algorithm detects and corrects distortions in extrema while enforcing that every data point maintains a consistent steepest ascending or descending direction, referred to as \emph{local ordering}, which involves only each data point and its directly adjacent neighbors without introducing global data dependencies, allowing efficient parallel execution, as shown in Figure~\ref{fig:msz}(b). Furthermore, the relaxed requirements for PLMSS preservation allow each block to perform corrections independently until local convergence. In summary, this paper makes the following contributions:

\begin{itemize}
    \item We propose a scalable distributed iterative algorithm for correcting PLMSS for error-bounded lossy compression by introducing a relaxed sufficient condition for PLMSS preservation that avoids global integral line tracing.
    \item We relax synchronization across iterations of our algorithm, which further decreases communication overhead.
    \item We demonstrated over 90\% parallel efficiency at 128 GPUs under weak scaling and provided a comprehensive evaluation across several scientific datasets, correcting topological errors from two state-of-the-art compressors (SZ3 and ZFP).
\end{itemize}

\section{Related Work}
We review the related work on compression for scientific data, topology-preserving lossy compression, and parallel integral path computation.
\subsection{Compression for Scientific Data}

Error-bounded lossy compression~\cite{sz, sz3, Tao_2017, Kai2021, Kai2020, Di2018, Di2016, huang2023cuszp, huang2024cuszp2, FPZIP, zfp, MGARD, Kai22} achieves high compression ratios while guaranteeing that the error remains within a user-specified bound. However, a key limitation of most existing compressors is that they only ensure pointwise accuracy without preserving features or topological structures that are critical for scientific analysis.
For a comprehensive review of error-bounded lossy compression for scientific datasets, we refer readers to Di et al.~\cite{Di2024}.

Error-bounded lossy compressors can be classified into prediction-based and transform-based methods based on how they decorrelate the data before quantization and encoding. The SZ series~\cite{sz, sz3, Tao_2017, Kai2021, Kai2020, Di2018, Di2016, huang2023cuszp, huang2024cuszp2} is a representative family of the prediction-based compressors that predict data values based on neighboring information, quantize the prediction residuals, and encode the residuals using techniques such as Huffman coding, followed by lossless compression with tools like ZSTD~\cite{zstd} or GZIP~\cite{gzip}. Recent works also explored using neural networks to improve prediction accuracy, including AE-SZ~\cite{liu_2021} and SRNN-SZ~\cite{liu2023srnsz}. Different from the SZ series, FPZIP~\cite{FPZIP} reduces data size by ignoring a specified number of bit planes while enabling on-demand control of data distortion. MGARD~\cite{MGARD} is another example of a transformation-based lossy compressor that applies wavelet theories to uniform/non-uniform structured and unstructured grids. 
ZFP~\cite{zfp} is a representative example of the transform-based method that transforms the data into sparse coefficients by applying a custom orthogonal block transform to decorrelate the data within blocks, which are then efficiently compressed. 

\subsection{Topology-Preserving Lossy Compression}\label{sec:feature_compression}
Topology-preserving lossy compression aims to preserve the topological feature of the data during compression, rather than merely ensuring pointwise accuracy. 
The topological descriptors, such as critical points and PLMSS, capture the structural organization of scalar fields and provide a high-level abstraction of complex data, forming the foundation for feature extraction, segmentation, and quantitative analysis in various scientific domains, including combustion, cosmology, and climate modeling~\cite{Yan_2021}.

While topology-preserving compression has been investigated for topological features~\cite{yan2023toposz, gorski2025general, Liang_2020, LiangDCRLOCPG23, Xia24, Xia25, li_msz}, existing methods are generally limited to shared-memory or single-node environments, restricting their applicability to large-scale datasets. For instance, MSz~\cite{li_msz} supports only single-GPU execution, and TopoSZ~\cite{yan2023toposz} lacks parallel design, restricting their applicability to large-scale datasets.

A widely adopted paradigm for topology-preserving compression is the \emph{edit-based} strategy, which performs topology correction after decompression when both the original and decompressed data are available in memory. The edit-based strategy identifies a subset of data points to be modified within the prescribed error bound to restore topological consistency. For example, MSz~\cite{li_msz} corrects distortions in the PLMSS through an iterative workflow by using the edit-based strategy. Our work builds on the edit-based strategy but extends it to a distributed-memory setting, allowing each data block to perform local corrections independently while maintaining global topological consistency, as described later.

As related, there exist alternative strategies for feature preservation, including modifying the input data prior to compression~\cite{soler2018topologically} and embedding feature constraints directly into the compression algorithm~\cite{yan2023toposz}. While effective, these approaches are often tightly coupled to specific compressors or require global data dependencies that limit their scalability. 

\subsection{Parallel Integral Path Computation}
Integral path computation is a fundamental operation in scientific visualization, used to trace trajectories such as streamlines and gradient integral lines~\cite{Peterka11}.
An integral path is a continuous curve whose tangent direction follows the underlying vector or gradient field, revealing flow structures or feature connectivity in the data.
In PLMSS, integral paths define segmentation boundaries by tracing from each data point along the gradient field to its corresponding extremum~\cite{MSS}.

However, efficiently parallelizing the computation of integral paths is challenging because each path exhibits strong sequential and non-local dependencies: each path's trajectory may traverse multiple data partitions, requiring irregular memory access and frequent inter-process communication in distributed environments.
Rather than optimizing this inherently difficult operation, our algorithm eliminates the need for integral path computation by relaxing the preservation requirements for PLMSS.

Existing parallel algorithms for integral path computation can be broadly divided into two categories: shared-memory and distributed-memory approaches.
Shared-memory approaches exploit thread-level parallelism within a single node to accelerate path tracing. Nouanesengsy et al.~\cite{Nouanesengsy11, Nouanesengsy12} and Müller et al.~\cite{Müller13} developed dynamic task scheduling and work-requesting schemes to balance irregular workloads during particle advection and Finite-Time Lyapunov Exponent computation.
Kendall et al.~\cite{Kendall11} and Peterka et al.~\cite{Peterka11} improved traversal and task migration strategies for large ensemble flow visualization.
Distributed-memory approaches extend integral path computation to large-scale environments where data are partitioned across processes.
Li and Fujishiro~\cite{Li08} and Garth et al.~\cite{Garth07} reduced inter-process communication via spatial decomposition and coherent structure detection.
Lu et al.~\cite{Kewei14}, Zhang et al.~\cite{Zhang16}, and Guo et al.~\cite{Guo13} improved scalability by optimizing data access and redistribution.
Morozov and Peterka~\cite{Morozov16} introduced DIY2 for block-parallel execution, and Will et al.~\cite{will2024} extended integral-line computation to large-scale Morse–Smale segmentation on distributed systems.

\section{BackGround}
We review the background of Morse-Smale Segmentation in Piecewise Linear Scalar Fields and the previous PLMSS preserving method (MSz~\cite{li_msz}), which is limited to a single GPU.

\subsection{Morse-Smale Segmentation in Piecewise Linear Scalar Fields}

\begin{figure}
    \includegraphics[width=\linewidth]{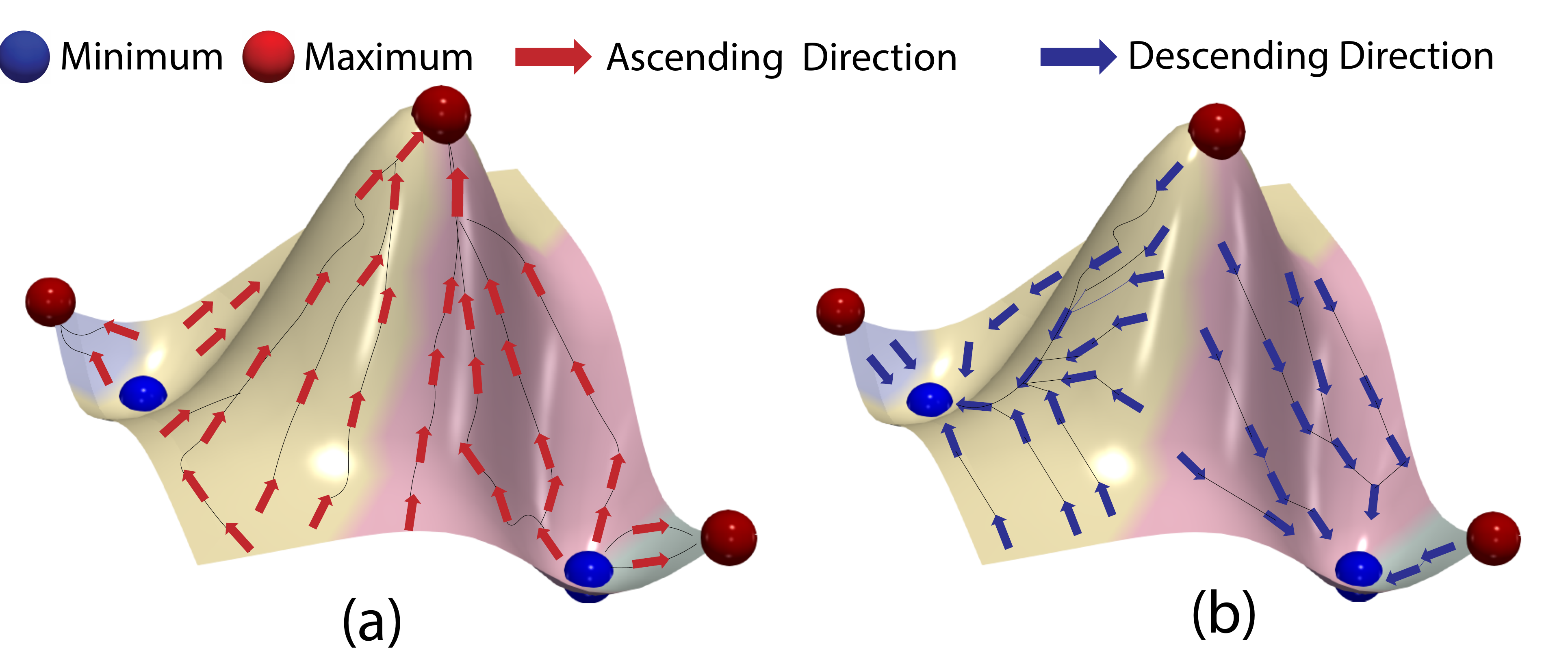}
    \caption{Illustration of MS segmentations. Red and blue spheres represent local maxima and minima. Black paths represent ascending and descending integral paths.}
    \label{fig:mss}
\end{figure}

We briefly review piecewise linear Morse-Smale segmentations (PLMSS)~\cite{MSS}. As illustrated in Figure~\ref{fig:mss}, for a given piecewise linear function (represented with a constant triangular/tetrahedral mesh) with a distinct value $f_i$ defined on data point $i$, PLMSS, as shown in Figure~\ref{fig:mss}, partitions the data into different regions of consistent gradient behavior: any point within the same region converges to the same minimum $m_j$ and maximum $M_k$ when traced along the steepest ascent and descent directions, as discussed below.

\noindent \textbf{Extrema} are data points whose scalar values are higher or lower than all their neighbors, shown as red (maxima) and blue (minima) spheres in Figure~\ref{fig:mss}.  
\textbf{Integral lines} are monotone paths along mesh edges, following the steepest ascent or descent direction (the red or blue arrows in Figure~\ref{fig:mss}) from each data point until reaching an extremum, as illustrated by the red and blue paths in Figure~\ref{fig:mss}(b).  
\textbf{Ascending} and \textbf{descending segmentations} group data points that converge to the same minimum or maximum, respectively.  
\noindent \textbf{Piecewise linear Morse–Smale segmentation (PLMSS)} combines the ascending and descending segmentations to produce a partition of the domain. Each region is defined by a pair of extrema $\langle m_j, M_k \rangle$ and separated from other regions by integral lines, as the regions of different colors shown in Figure~\ref{fig:mss}.

\subsection{Previous work on single-GPU correction of PLMSS}~\label{sec:MSz}
The previous work MSz~\cite{li_msz} is a single-GPU algorithm that preserves PLMSS in error-bounded lossy compression by iteratively correcting topological distortions in the decompressed data. However, MSz relies on explicit integral path computation, which introduces strong data dependencies and limits its scalability in distributed-memory environments.

A key difference between our algorithm and MSz is that our method eliminates explicit integral path computation, which is the primary scalability bottleneck of MSz, thus improving its efficiency, particularly in distributed-memory environments. To better illustrate the difference, we briefly review the design of MSz below.  

MSz takes as input a 2D or 3D scalar field $f$ and its decompressed version $\hat{f}$ obtained from an arbitrary error-bounded lossy compressor with a global error bound $\xi$. 
The algorithm outputs a set of edits $\delta = \{\delta_i\}$ applied to $\hat{f}$ to obtain a corrected field $g = \hat{f} + \delta$ that preserves the PLMSS and strictly satisfies the global error bound $|f_i - g_i| \leq \xi$ for all data points.

The key idea of MSz is the edit-based strategy that identifies a subset of data points in the decompressed data whose values can be modified within the user-specified error bound to correct topological distortions. In error-bounded lossy compression, topological distortions primarily arise because compression errors perturb the scalar ordering among neighboring data points, which is fundamental to the topology of the underlying scalar field. The edit-based strategy restores the necessary ordering by adjusting selected data values within the error bound to correct the distorted topology.

Specifically, the edit-based strategy iteratively detects topology distortions, identifies the data point $i$ responsible for the distortion, and applies edits $\delta_i$ to restore the topological consistency.
Let $f_i$ denote $i$'s original value, $g_i^{(k)}$ the modified value at iteration $k$, and $\xi$ is the user-specified error bound. Each edit updates the scalar value of $i$ according to the following equation:
\begin{equation}
g_{i}^{(k+1)} = 
    \begin{cases}
    g_{i}^{(k)} - \delta_i, & g_{i}^{(k)} - \delta_i \geq f_i - \xi, \\
    f_i - \xi, & \text{otherwise}.
    \end{cases}
\label{eq1}
\end{equation}

MSz consists of two alternative iterative subloops built upon the edit-based strategy to preserve PLMSS in error-bounded lossy compressed data: the C-loops and the R-loops, as discussed below.

\noindent \textbf{C-Loops} focuses on preserving local extrema in the decompressed data, as the local extrema are fundamental constituents for PLMSS. MSz corrects four types of distortion in the c-loops: false positive maxima (FPmax), false negative maxima (FNmax), false positive minima (FPmin), and false negative minima (FNmin). We use the FPmax case to illustrate the rule to identify the data point $i$ to be modified.
\begin{figure}[htb!]
    \centering
    \includegraphics[width=\linewidth]{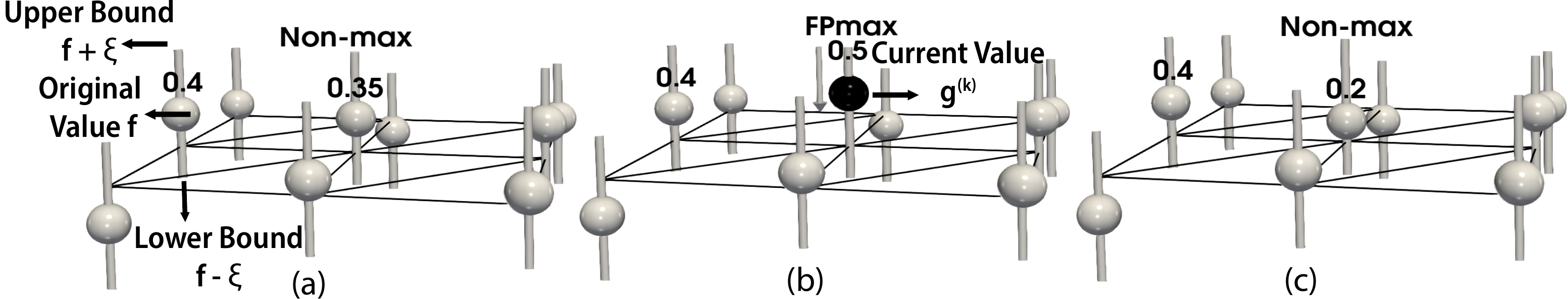}
    \caption{Illustration of the correction process for false-positive maxima. The vertical position of each sphere shows its scalar value at the current iteration. Black spheres indicate the distortion, white spheres are non-extrema, and cylinders represent the value range of each data point $[f-\xi, f+\xi]$.}
    \label{fig:fpmax}
    
\end{figure}

Specifically, a data point $i$ is considered an FPmax if there exists at least one neighboring point $j$ whose scalar value is higher than $f_i$, such that $f_j > f_i$ in the original data $f$ (as the sphere with a scalar value of 0.35 in Figure~\ref{fig:fpmax} (a)), but no such neighbor exists in the decompressed data $\hat{f}$, causing $i$ to appear as a maximum incorrectly, as the black sphere shown in Figure~\ref{fig:fpmax} (b).  
In this case, we decrease $g_i$ by $\delta_{\text{edit}}$ to restore the correct ordering, as shown in Figure~\ref{fig:fpmax}.
For the remaining cases, we refer readers to MSz~\cite{li_msz} for details of a similar correction strategy.

\noindent\textbf{R-Loops.} Once all extrema are preserved, MSz proceeds to the R-loop, which aims to correct wrongly-labeled data points by repeatedly tracing integral lines to locate and edit the \emph{troublemaker} (the first occurrence of discrepancy along the integral line). 
The R-loop is extremely computationally expensive and non-trivial to efficiently parallelize because each iteration requires recomputing integral paths and becomes the primary bottleneck in MSz. In distributed-memory environments, as in our naive parallelization of MSz, the cost is further increased since each integral path may traverse multiple data blocks, requiring inter-process communication to access data required for path tracing.
In contrast, our algorithm completely eliminates the need for the R-loop, ensuring PLMSS consistency without performing any integral path computation.

\noindent\textbf{Convergence of the iterative strategy.} Note that each modification during the iterative process may introduce new distortions, requiring further corrections in subsequent iterations. Nevertheless, convergence is guaranteed. The key rationale is that each edit applied to a data point $i$ monotonically decreases its value toward the lower bound $f_i - \xi$, i.e.,$f_i = g_i^{(0)} \geq g_i^{(1)} \geq g_i^{(2)} \geq \cdots \geq g_i^{(k)} > f_i - \xi$. Therefore, no data point can be updated indefinitely. In the worst case, every data point $i$ converges to its lower bound $f_i - \xi$, at which point the preserved extrema and local ordering remain consistent with the original field, ensuring that the PLMSS of the corrected field $g$ matches that of $f$, guaranteeing the convergence of the iterative process.  For a detailed proof of convergence, we refer readers to MSz~\cite{li_msz}.

\noindent\textbf {Scalability Limitations of MSz.}
While MSz effectively preserves the PLMSS on a single GPU, its design introduces significant scalability challenges when extended to distributed-memory environments for two primary reasons.

First, the R-loop requires recomputing integral paths in every iteration, introducing strong global data dependencies.
Each integral path may traverse large regions of the domain and cross multiple process boundaries, resulting in frequent inter-process communication during path tracing.
Second, additional data communication is required in every iteration to synchronize the updated integral path results across neighboring blocks, further increasing the overall cost.
Our method addresses these limitations by removing the R-loop and redesigning the C-loop, as discussed in the following section.

\section{Our Method}~\label{sec:method}
To address the scalability limitations of MSz~\cite{li_msz}, our algorithm removes the computationally expensive R-loop and redesigns the C-loop to ensure fully local corrections with efficient synchronization across data blocks. 
Specifically, we (1) simplified MSz by eliminating the global dependencies introduced by integral path tracing, removing the R-loop and modifying the C-loop to preserve PLMSS through purely local operations, and (2) enabled scalable distributed parallelization through blockwise processing, ghost-layer synchronization, and relaxed interprocess communication frequency.

The rest of this section presents our distributed algorithm for preserving PLMSS under error-bounded lossy compression from two perspectives: (1) the simplification of MSz (Section~\ref{sec:simplification}), (2) the distributed parallel design that ensures scalability and efficient communication (Section~\ref{sec:parallism}), and (3) the convergence analysis of our iterative correction process in the distributed-memory environment (Section~\ref{sec:Convergence}).

Notations used throughout our paper are as follows: $\xi$ represents the user-defined error bound, $i$ denotes the $i$th data point. The scalar values at $i$ in the original and decompressed data are represented by $f_i$ and $\hat{f}_i$, respectively, while $g_i$ refers to the edited scalar value at $i$. 
\begin{figure*}[htb]
    \centering
    \includegraphics[width=\linewidth]{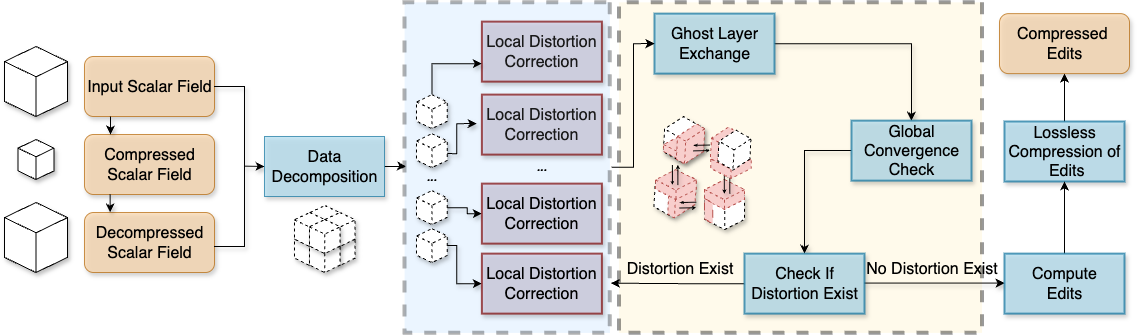}
    \caption{Overview of the iterative workflow. The input is the decompressed data from an error-bounded lossy compressor, and the output is a set of edits. The field is partitioned into blocks with ghost layers, and each iteration performs local distortion correction, ghost layer synchronization, and convergence checking.}
    \label{fig:workflow}
\end{figure*}
\subsection{Simplification of MSz: Removing the Computationally Intensive R-Loops}~\label{sec:simplification}
Our algorithm eliminates the R-loop entirely and instead enforces a relaxed sufficient condition for PLMSS preservation. Specifically, rather than explicitly tracing integral paths to determine correct segmentations, we ensure that (1) all extrema are preserved, and (2) the local ordering of each data point, its smallest and largest neighbors, remains consistent with the original field. The relaxed conditions for PLMSS preservation are sufficient to guarantee that each data point follows the same gradient direction toward its corresponding extremum (i.e., the arrows in Figure~\ref{fig:mss}), thus preserving the PLMSS without computing integral paths.

While preserving the local ordering of all data points is sufficient for maintaining the same integral paths, this relaxed condition may introduce more edits compared to MSz.
However, as demonstrated in our evaluation, the number of additional edits is very small (as shown in Figure~\ref{fig:msz_comparison}), whereas the relaxed formulation yields substantial performance gains by eliminating R-loops and enabling fully local corrections.

\begin{figure}[htb!]
    \centering
    \includegraphics[width=\linewidth]{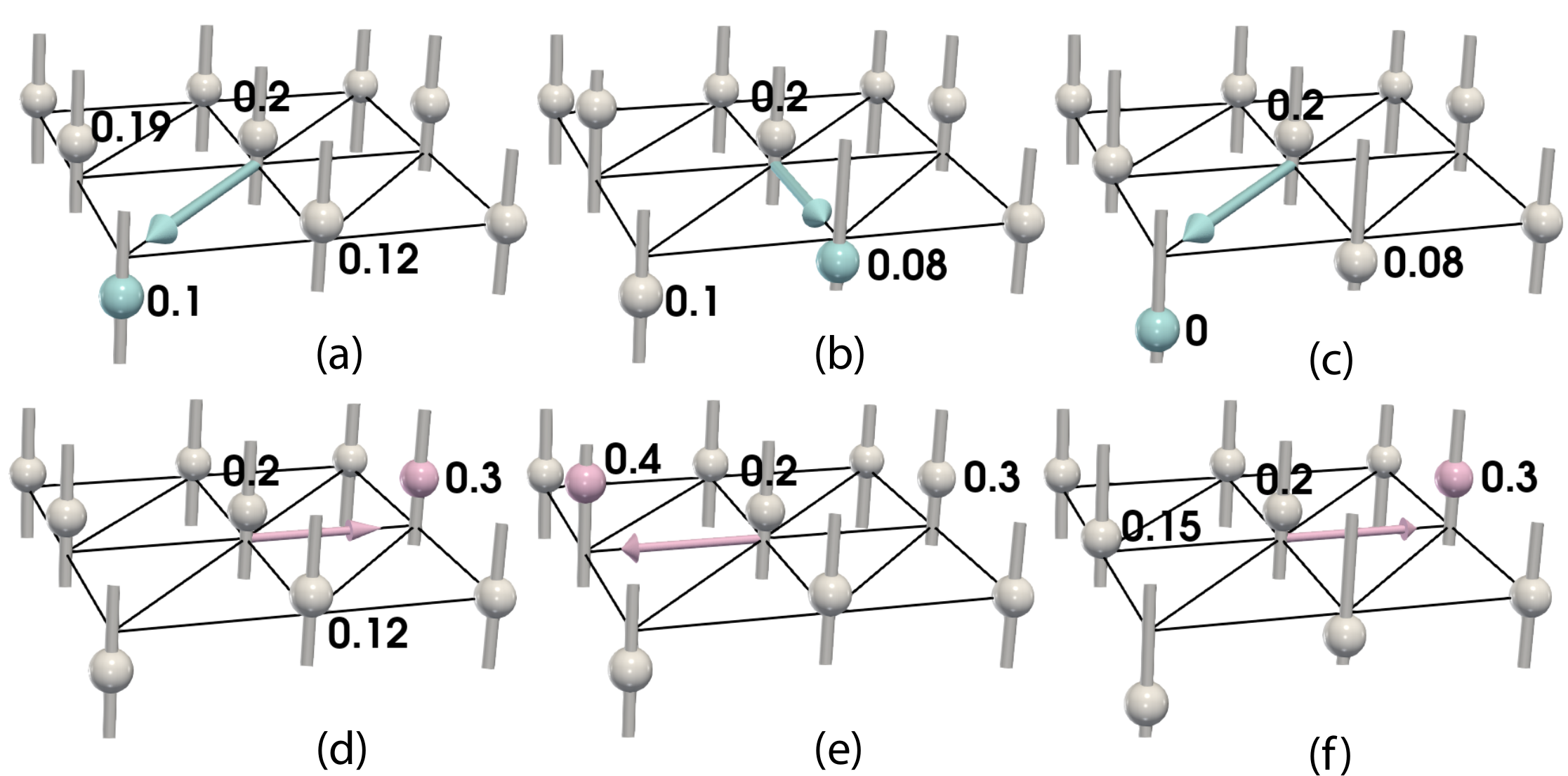}
    \caption{Illustration of the correction process for local ordering.
(a–c) show the correction steps for the descending direction, while (d–f) show those for the ascending direction.
Blue and pink arrows indicate descending and ascending directions, respectively, and the blue/pink spheres denote the smallest/largest neighbors of the central sphere in each iteration.}
    \label{fig:nonextrema}
\end{figure}
Specifically, for the preservation of extrema, we follow the same rule as in the C-loop of MSz~\cite{li_msz}, ensuring that all local minima and maxima remain consistent with the original data.
For local ordering preservation, we redesigned the rules in the C-loop of MSz to enforce the consistency of the local ordering between each data point and its neighborhood: each data point $i$, we require that the speediest ascending and descending direction of each data point remain the same as in the original data $f$, as discussed below.

For each data point $i$, let $n_{\min}(i)$ and $n_{\max}(i)$ denote its smallest and largest neighbors in the original data $f$, i.e., 
$n_{\min}(i) = \arg\min_{j \in \mathcal{N}(i)} f_j$ and $n_{\max}(i) = \arg\max_{j \in \mathcal{N}(i)} f_j$. 
As shown in Figure~\ref{fig:nonextrema}, the blue and pink spheres represent $n_{\min}(i)$ and $n_{\max}(i)$, respectively. 
In the decompressed data $\hat{f}$, the corresponding neighbors are denoted by $\hat{n}_{\min}(i)$ and $\hat{n}_{\max}(i)$.

\noindent\textbf{Descending direction.}  
In $f$ (Figure~\ref{fig:nonextrema}(a)), the smallest neighbor of $i$ is the blue sphere with value $0.1$, i.e., $n_{\min}(i)$ has $f_{n_{\min}(i)}=0.1$.  
After compression (Figure~\ref{fig:nonextrema}(b)), the local ordering is distorted: a different neighbor becomes the smallest in $g$ with value $0.08$, so $\hat n_{\min}(i)\neq n_{\min}(i)$.  
Because our edits are non-increasing, we cannot increase the data point with value $0.08$; instead, we decrease the value at the original minimum so that it again becomes the smallest among all neighbors while obeying the lower bound.  
We decrease $g_{n_{\min}(i)}$ from $0.1$ to $0$ (Figure~\ref{fig:nonextrema}(c)), which restores the correct descending relation around $i$ (the neighbor along the descending direction is once again the original $n_{\min}(i)$).
\noindent\textbf{Ascending direction.}  
In $f$ (Figure~\ref{fig:nonextrema}(d)), the pink sphere with value $0.3$ is the largest neighbor of $i$ (i.e., $n_{\max}(i)$).  
After compression (Figure~\ref{fig:nonextrema}(e)), another neighbor becomes spuriously larger with value $0.4$, so $\hat n_{\max}(i)\neq n_{\max}(i)$.  
Here we can directly \emph{decrease the spurious maximum}: we reduce $g_{\hat n_{\max}(i)}$ from $0.4$ to $0.3$ (or slightly below within tolerance), yielding the ordering shown in Figure~\ref{fig:nonextrema}(f), e.g., along the ascending direction the values follow $0.15 \rightarrow 0.2 \rightarrow 0.3$, consistent with the original $f$.

In both cases, edits are monotonic (non-increasing) and bounded by the local tolerance, ensuring no new extrema are introduced while the local ordering around $i$ matches that of the original field.

\subsection{Distributed Parallel Algorithm}~\label{sec:parallism}

\begin{figure}
    \centering
    \includegraphics[width=\linewidth]{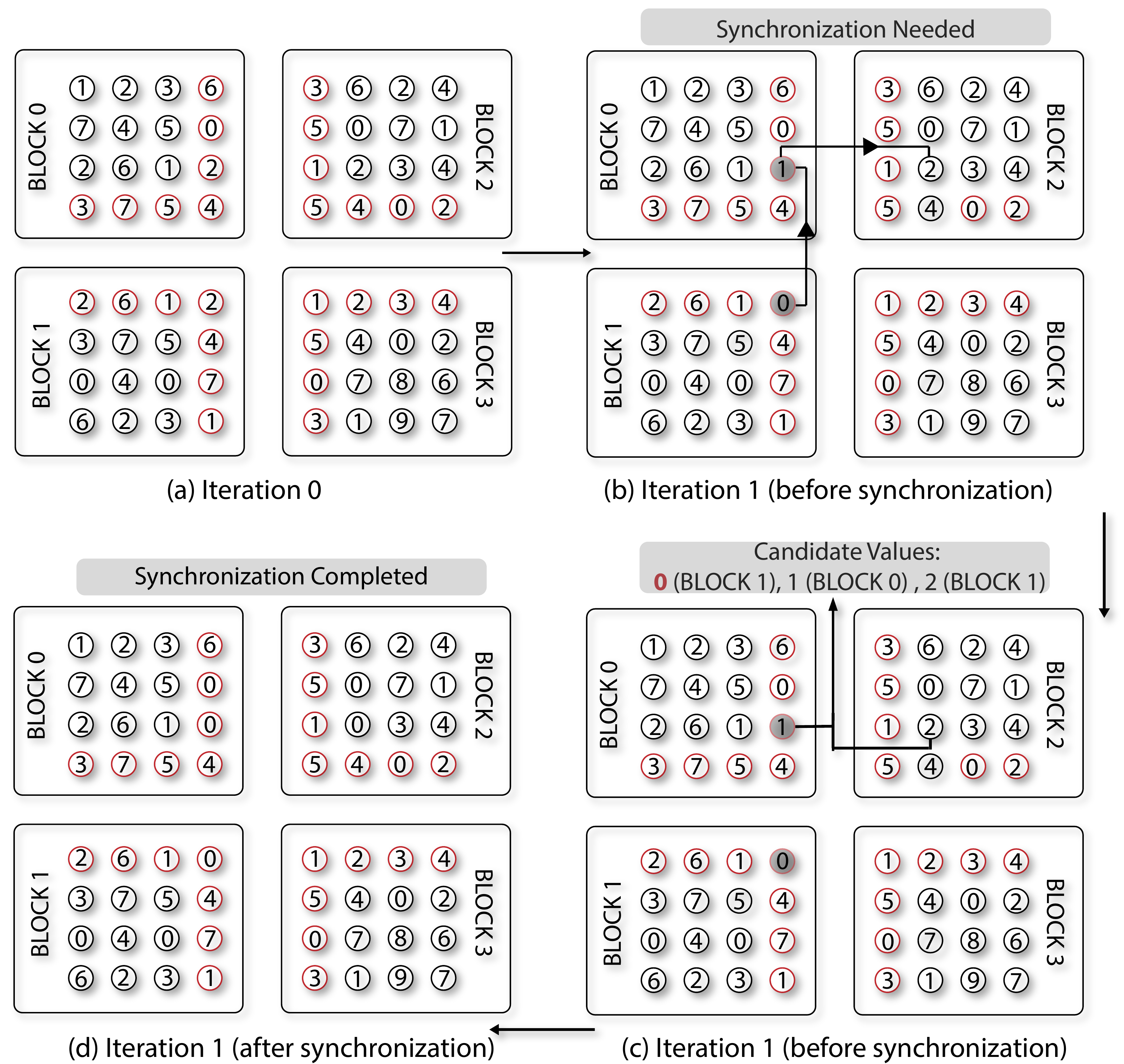}
    \caption{Each data block is associated with a red-bordered ghost layer that provides neighborhood information across block boundaries.
Gray points indicate data values that have been modified during local correction.}
    \label{fig:ghost}
\end{figure}

To scale our method to large-scale datasets, we design a distributed parallel algorithm that performs local corrections independently within each data block while maintaining consistency across block boundaries.
The key idea is to perform local corrections independently within each data block while ensuring global consistency by exchanging ghost layer values through lightweight communication across block boundaries. 
Our distributed algorithm consists of three main components: domain decomposition and local correction, and ghost layer synchronization.
\noindent\textbf{Domain Decomposition and Local Correction.}  
We partition the data domain into non-overlapping regions, each extended by a single-layer ghost region to exchange boundary information, each assigned to a GPU, as shown in Figure~\ref{fig:ghost}, where the red-bordered points represent ghost regions.
Each block performs local correction independently, following the relaxed preservation condition introduced earlier, ensuring that all extrema and local orderings are preserved within the block.

\noindent\textbf{Ghost Layer Synchronization.}  
After each block completes local correction, neighboring blocks exchange ghost layers to ensure consistent scalar values along shared boundaries.
When multiple blocks update the same boundary point $i$ with different values, conflicts are resolved deterministically by selecting the minimum among all candidate values, which guarantees monotonic non-increasing updates toward the lower error bound $f_i - \xi$. As illustrated in Figure~\ref{fig:ghost}, both Block~0 and Block~1 modify the same boundary point (the gray point in Figure~\ref{fig:ghost} (b)) with values~1 and~2, respectively, while Block~2 retains the original value~0.
During synchronization, these three candidate values (0, 1, 2) are reconciled by selecting the minimum value~0, as shown in Figure~\ref{fig:ghost} (d), ensuring consistent and monotonic updates across neighboring blocks.

\begin{figure}[htb!]
    \centering
    \includegraphics[width=\linewidth]{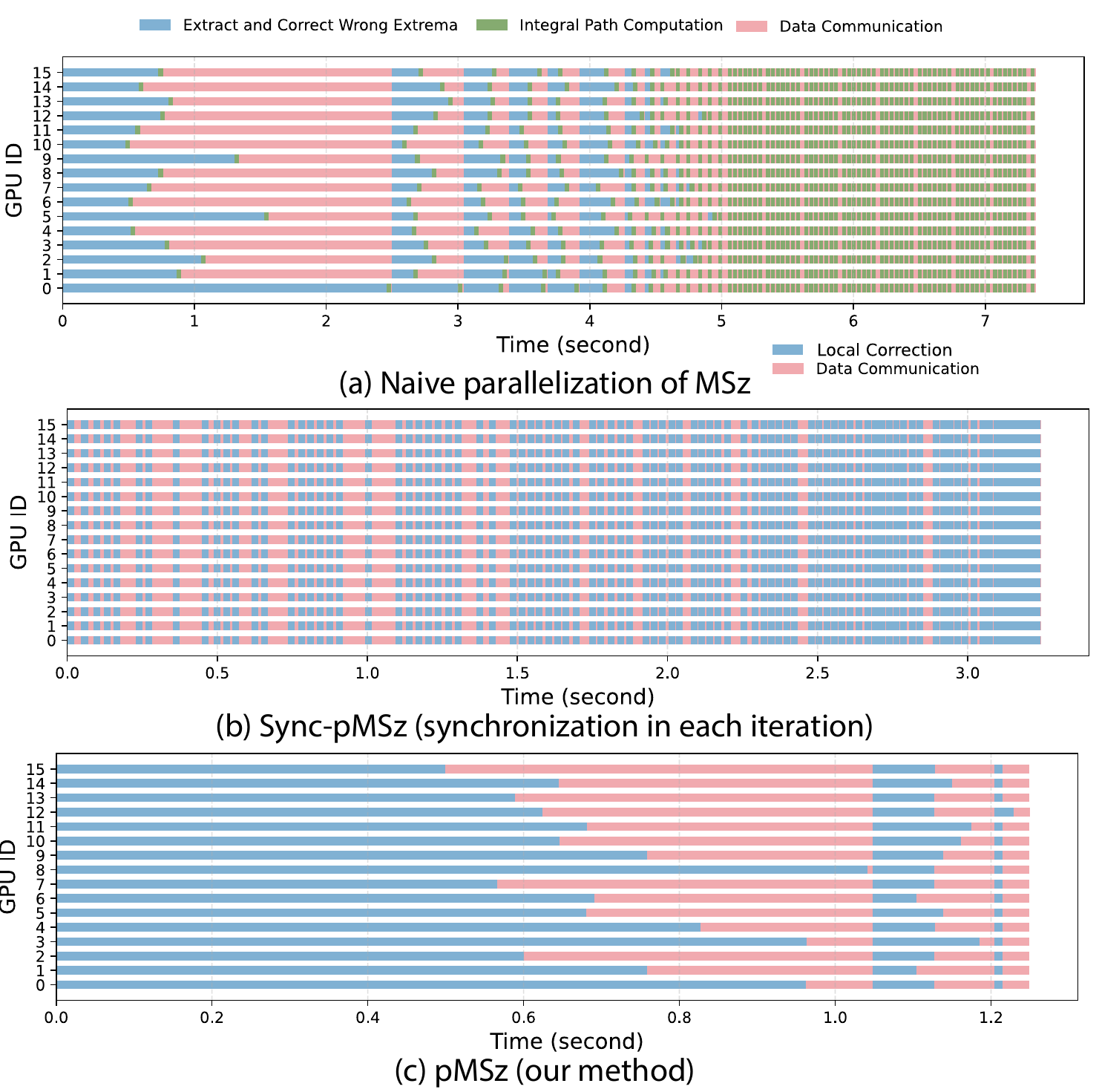}
    \caption{Comparison of computation (blue and green) and communication/waiting (pink) phases across iterations for (a) naive parallelization of MSz, (b) Sync-pMSz (a baseline with per-iteration synchronization of our method), and (c) pMSz (our method). }
    \label{fig:gantt}
\end{figure}

\vspace{0.3em}
\noindent\textbf{Relaxed Synchronization.}
Although ghost layer synchronization maintains correctness across block boundaries, its frequency has a direct impact on overall performance.
Two synchronization strategies are possible:
(1) Per-Iteration Synchronization: updating ghost data points immediately after each local iteration ensures immediate consistency but leads to frequent communication, as each iteration incurs a synchronization phase, as shown in Figure~\ref{fig:gantt} (b) 
or
(2) performing synchronization only after the local correction within a block has converged, as shown in Figure~\ref{fig:gantt} (c).
We adopt the second strategy because corrections to core data points are confined within local blocks, and only boundary points depend on neighboring data.
By allowing each block to reach local convergence before synchronization, unnecessary communication is avoided while maintaining correctness along block interfaces.
Our relaxed synchronization introduces a trade-off: it may require more local correction iterations to reach global convergence, because boundary updates are deferred.
However, in practice, the frequency of ghost layer synchronization is greatly reduced, as shown in Figure~\ref{fig:gantt}, resulting in substantial performance gains.

\subsection{Convergence Analysis}\label{sec:Convergence}
Our algorithm alternates between local correction and ghost layer synchronization until all data blocks satisfy the relaxed condition for PLMSS preservation.  
The convergence of the local correction within each block has already been proved, as discussed in Section~\ref{sec:MSz}, and our method inherits this property because the same edit rules are applied locally under the global error bound $|f_i-\hat f_i|\le\xi$.
However, after local correction converges, neighboring blocks may still hold different scalar values for shared boundary points, as illustrated in Figure~\ref{fig:ghost}. The inconsistencies in the boundary points could introduce new local distortions near block boundaries, which will vanish with a finite number of iterations with our decreasing edit strategies.
In the ghost layer synchronization step, each shared ghost data point takes the minimum of all candidate values from its neighboring blocks, ensuring that updates across blocks remain non-increasing and that all values stay within the global error bound. Because synchronization only decreases values and never violates the error bound, our algorithm is guaranteed to converge.

As a result, both local correction and synchronization produce updates that are non-increasing and bounded by $f_i-\xi$.  
Theoretically, in the most extreme case, all data points (including those in the ghost layers) reach their lower bounds, at which point the local ordering is fully restored.  
Therefore, the distributed iterative process converges to a stable field $g$ that satisfies both the global error bound and the relaxed sufficient condition for PLMSS preservation.

\section{Evaluation}
In this section, we evaluate our method from three perspectives: 
(1) the effectiveness and trade-offs of the algorithmic simplification that eliminates the R-Loops in MSz, and 
(2) the scalability and performance of our distributed parallel design.
(3) the ability to preserve PLMSS of our method.

The experiments highlight three major findings:  
\begin{itemize}
    \item \textbf{Scalability.} Our method scales to 128 GPUs under weak scaling with parallel efficiency above 90\%, whereas the naive parallelization of MSz only achieves 8.66\%.  
    \item \textbf{Single-GPU Performance.} Compared with MSz, our method achieves up to $14\times$ speedup on a single GPU by eliminating the R-loop and reformulating the preservation condition to enable fully local corrections.
    \item \textbf{Compression effectiveness.} Our method achieves up to $88\times$ compression ratios under 100\% PLMSS preservation across the tested datasets, whereas the base compressors can only preserve PLMSS in lossless mode or at extremely low compression ratios.
\end{itemize}

\subsection{Experimental Setup} 
All experiments are performed on NERSC's Perlmutter system, which is equipped with 4 NVIDIA A100 GPUs (40 GB memory per GPU) per compute node. Our implementation is developed in CUDA and CUDA-aware MPI, enabling direct use of GPU memory for data communication. 
For scalability studies, we employ up to 128 GPUs with weak scaling, where each rank processes a fixed $512^3$ subdomain.
We first summarize the baselines used in our experiments before introducing the details of our method and evaluation.
\noindent\textbf{Baseline~1: Single-GPU MSz.}
This baseline uses the original MSz~\cite{li_msz} implemented on a single GPU.
It serves as the reference for evaluating our algorithmic simplification, particularly the removal of the R-loops.
\noindent\textbf{Baseline~2: naive-MSz.}
To evaluate scalability improvements, we implemented a distributed-memory version of MSz by partitioning the domain across multiple GPUs.
Each GPU independently executes the original MSz algorithm, and ghost regions are exchanged after every R-loop iteration.
\noindent\textbf{Baseline~3: sync-pMSz.}
This baseline uses our method but performs synchronization across neighboring blocks after \emph{every iteration} and is used to analyze the impact of synchronization frequency in our relaxed communication strategy.

For single-GPU evaluations, we use the NYX dataset to compare our simplified formulation with the original MSz.  
For scalability studies, we use synthetic Perlin noise datasets of increasing global sizes to ensure consistent data characteristics across scales.  
For PLMSS preservation evaluation, we use all real-world datasets from cosmology, combustion, molecular dynamics, and climate simulations to test the compression effectiveness, overhead, and topological correctness of our method. 

\subsection{Single-GPU Evaluation: Effectiveness of Simplification}

This section evaluates the effectiveness of our algorithmic simplification that removes the R-loop in MSz while preserving PLMSS under error-bounded lossy compression on the NYX dataset using SZ3 with a relative error bound of $10^{-6}$.
The single-GPU experiments compare pMSz (our method) with the original MSz to demonstrate the benefit of removing global integral path computations.

\begin{figure}[htb!]
    \centering
    \includegraphics[width=\linewidth]{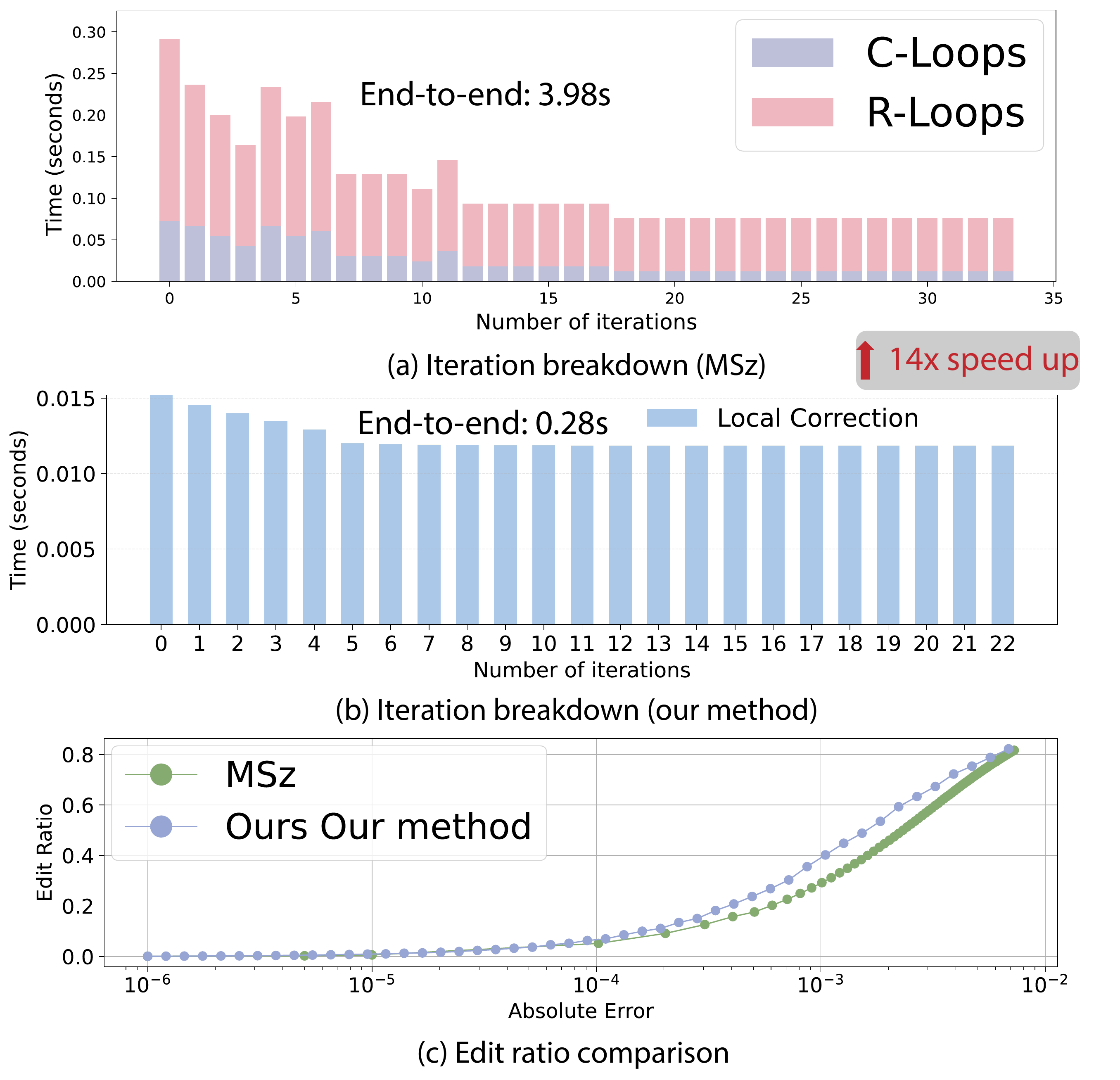}
    \caption{Timings of iterations for (a) MSz and (b) our method on a single GPU, and (c) comparison of the edit ratio (modified points over total points) between MSz and our method.}
    \label{fig:msz_comparison}
    \vspace{-0.1in}
\end{figure}

As shown in Figure~\ref{fig:msz_comparison} (b), our method achieves up to $14\times$ speedup over MSz, with a slight increase in the number of data points that need modification, as shown in Figure~\ref{fig:msz_comparison} (c). 
This performance gain primarily comes from eliminating the R-loop, which dominates MSz's runtime due to its repeated global integral path computations for correcting mislabeled data points, as shown in Figure~\ref{fig:msz_comparison}(a), R-loops account for more than 80\% of the total runtime in later iterations in MSz, whereas the cost of C-loops becomes negligible. 

\subsection{Multi-GPU Evaluation: Scalability and Communication Efficiency}
The multi-GPU evaluation focuses on the distributed scalability and communication efficiency of our method.
We compare our distributed algorithm against two multi-GPU baselines: 
(1) a naive parallelization of MSz and (2) Sync-pMSz.
\subsubsection{Comparison with naive-MSz}
To illustrate the scalability challenge of MSz, we implemented a naive distributed version that partitions the data across multiple GPUs.
Each GPU executes the original MSz locally and exchanges ghost regions after every R-loop.
As shown in Figure~\ref{fig:gantt} (a), R-loops dominate the total runtime and involve intensive data communication across processes, accounting for up to 60\% of computation time.
Even compared with a baseline that performs synchronization after every iteration, the naive parallel MSz remains considerably slower due to repeated global integral path computations.
In contrast, our method avoids R-loops and allows each block to iteratively correct distortions until local convergence before synchronization, reducing communication frequency.

Compared with sync-pMSz, pMSz introduces a slight workload imbalance, as different blocks may require different numbers of iterations for local convergence (Figure~\ref{fig:ghost}(b)).
While this imbalance is minor, it can lead to idle time for blocks that finish earlier; exploring asynchronous communication strategies is left as future work.

\subsubsection{Scalability Study}
We next evaluate the scalability of our distributed algorithm under both weak and strong scaling to assess its parallel efficiency on large-scale GPU systems.

\noindent\textbf{Weak Scaling.}
We use the synthetic datasets generated with Perlin noise to evaluate the weak scalability of our algorithm. Each GPU processes a fixed $512^3$ subdomain, and the number of GPUs increases up to 128, resulting in a global domain of $2048 \times 2048 \times 4096$. 
Although each GPU handles the same data size, the overall problem complexity differs because both different datasets and different spatial blocks within the same dataset exhibit varying distortion levels, as shown in Figure~\ref{fig:weak_scaling} (a) and (c). Blocks that contain regions with more severe distortions require more local corrections and thus more iterations to reach convergence, as illustrated in Figure~\ref{fig:gantt}(c), where different blocks exhibit different local correction times. 
To eliminate this variability, we evaluate weak scaling using the average computation time per iteration, which isolates the scalability of each iteration from the dataset-dependent convergence behavior and provides a fair measure of parallel efficiency across different problem sizes.
As shown in Figure~\ref{fig:weak_scaling} (b), our algorithm maintains stable performance under weak scaling, achieving over 90\% parallel efficiency at 128 GPUs, where the parallel efficiency is computed as $E_p = T_1 / T_p$, where $T_1$ and $T_p$ denote the per-iteration runtimes on one and $p$ GPUs, respectively.

\begin{figure}
    \centering
    \includegraphics[width=\linewidth]{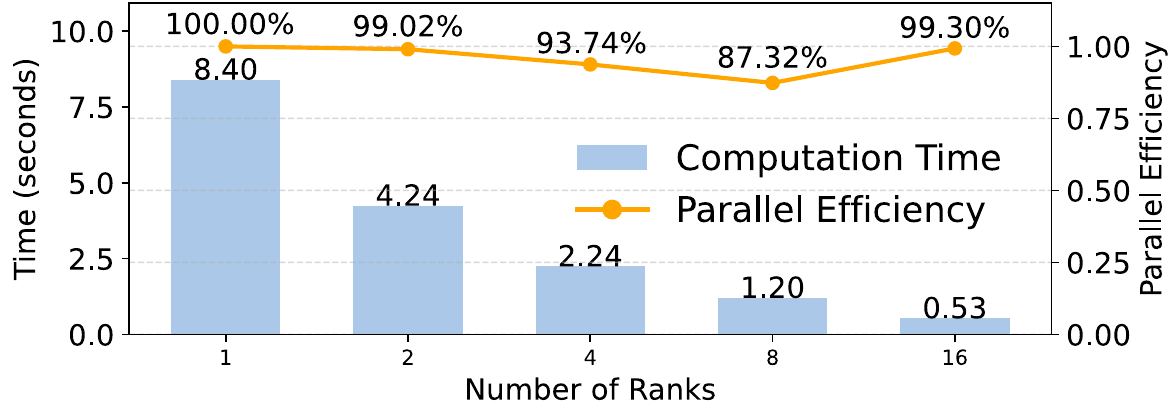}
    \caption{Strong scaling results of our method on a fixed-size $1024^3$ Perlin noise dataset.  
    Bars show computation time, and the line shows the corresponding parallel efficiency.}
    \label{fig:strong_scaling}
\end{figure}

\noindent\textbf{Strong Scaling.}
We further evaluate strong scaling using a fixed $1024^3$ Perlin noise dataset, increasing GPUs from 1 to 16.
Here, we evaluate the end-to-end runtime instead of per-iteration time, because all cases process the same global dataset, and the overall problem size (the PLMSS distortion) remains identical across different numbers of GPUs
As shown in Figure~\ref{fig:strong_scaling}, our method sustains high parallel efficiency across all tested configurations, remaining above 87\% and reaching 99.3\% at 16 GPUs, where the efficiency is computed as $E_p = T_1 / (p \cdot T_p)$.

\subsubsection{Impact of Relaxed Synchronization Strategy}
\begin{figure}
    \centering
    \includegraphics[width=\linewidth]{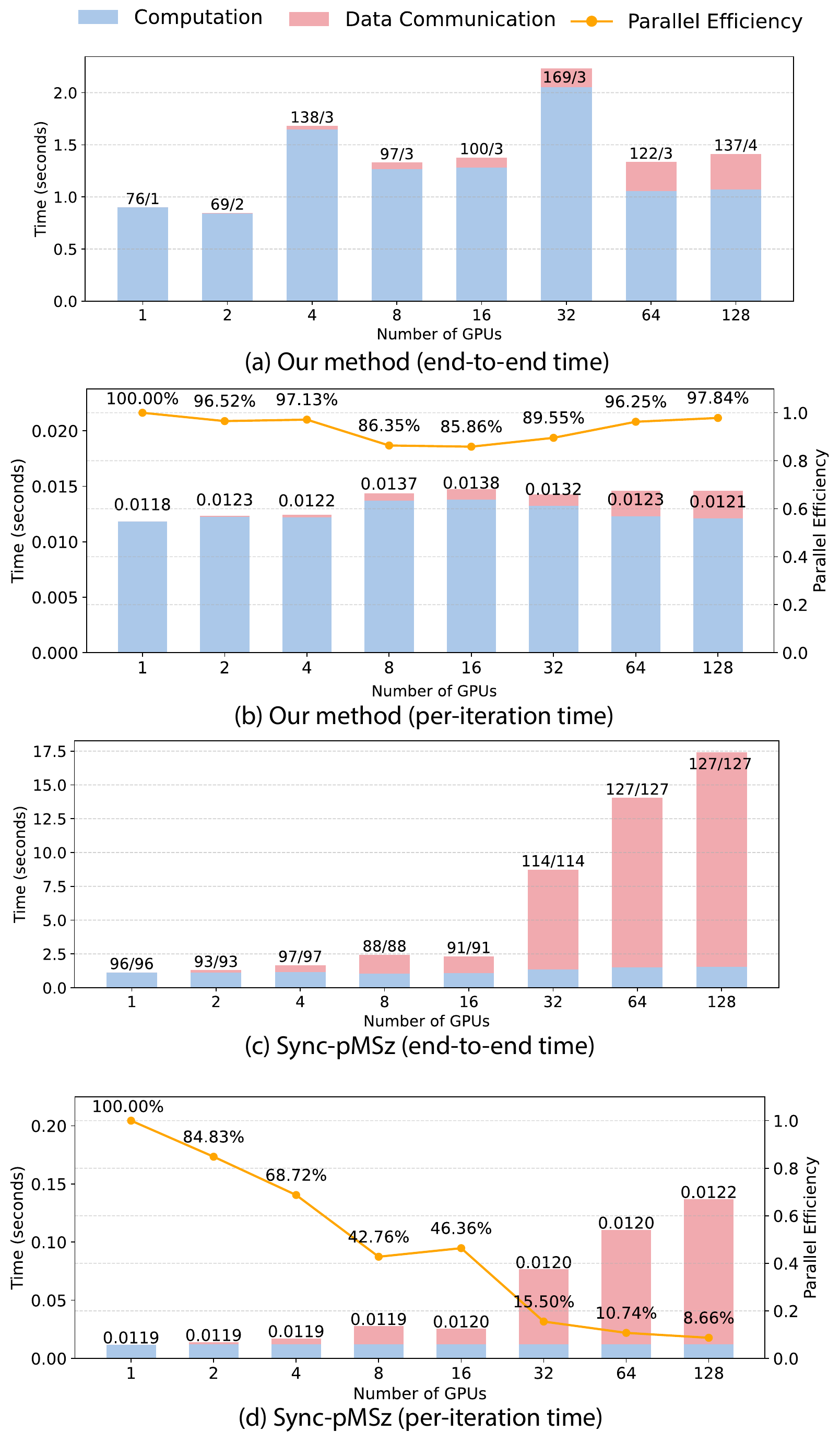}
    \caption{
Weak scaling comparison between our method (a,b) and a baseline with communication at every iteration (c,d). 
(a,c) Runtime breakdown into computation and communication. 
(b,d) Average per-iteration computation time (bars) and parallel efficiency (lines).
The numbers above each bar represent: (a,c) the number of iterations/number of synchronizations, and (b,d) the average per-iteration computation time.
}
    \label{fig:weak_scaling}
    \vspace{-0.1in}
\end{figure}

We analyze the effect of synchronization frequency by comparing pMSz (our method) with sync-pMSz.
As shown in Figure~\ref{fig:weak_scaling}(c,d), our method sustains up to 97\% parallel efficiency at 128 GPUs, whereas sync-pMSz drops to only 8\%.
By allowing each block to reach local convergence before communication, our method reduces synchronization frequency by more than 30× (at most three global exchanges vs. over 127 in the baseline), keeping communication lightweight and scalable.

\subsection{PLMSS Preservation Evaluation}
To evaluate the ability of our method to preserve PLMSS, we conduct experiments to analyze (1) the achievable maximum compression ratios under PLMSS preservation, (2) the computation and storage overhead introduced by additional edits, and (3) the effectiveness in preserving PLMSS across datasets. 

\noindent\textbf{Maximum Compression Ratio under PLMSS Preservation}
We evaluate the maximum compression ratios (CRs) achievable under PLMSS preservation across different datasets and base compressors (SZ3 and ZFP), as shown in Table~\ref{tab:plmss_cr}. Base compressors can typically preserve PLMSS only at lossless or very low compression ratios. In contrast, our method achieves up to 88$\times$ compression under PLMSS preservation, representing up to a 75$\times$ improvement over the base compressors across the tested datasets.

In addition, the maximum CRs vary across datasets due to differences in topological complexity. Datasets with more intricate PLMSS or higher sensitivity to perturbations tend to restrict achievable compression. For example, the Turbulent Jet dataset contains a large fraction of padded values, which amplifies PLMSS distortions and leads to relatively lower CRs compared with other datasets when considering the preservation of PLMSS. Another overall trend is that ZFP (ours) consistently achieves lower CRs than SZ3 (ours) under PLMSS preservation, since the raw compression ratios of ZFP are lower than those of SZ3.
\begin{table}[htb!]
\centering
\caption{Maximum compression ratio (CR) achieved by different compressors under PLMSS preservation. Values marked with $^*$ indicate that only lossless compression can preserve PLMSS.}
\label{tab:plmss_cr}
\renewcommand{\arraystretch}{1.2}
\setlength{\tabcolsep}{1pt}
\begin{tabular}{lccccc}
\hline
\textbf{Dataset} &\textbf{Size}& \textbf{SZ3} & \textbf{SZ3 with edits} & \textbf{ZFP} & \textbf{ZFP with edits} \\ 
\hline
NYX (dark density) & 8GB  &  1.20$\times^*$  &   \textbf{8.24$\times$}   &    1.01$\times^*$   & \textbf{7.35$\times$} \\
NYX          & 1GB  & 1.06$\times$   &   \textbf{8.49$\times$}  &   1.16$\times^*$ & \textbf{7.78$\times$}    \\
Turbulent Jet          & 317MB  &   1.96$\times$   &  \textbf{3.90$\times$}  &   1.70$\times$   & \textbf{4.42$\times$} \\
CSEM        & 50MB   &   1.06$\times^*$  &  \textbf{10.75$\times$} &   1.07$\times^*$   &  \textbf{2.45$\times$} \\
Vortex       &16MB  &  1.93$\times$   &   \textbf{25.80$\times$}  &   2.80$\times$    &    \textbf{11.48$\times$}  \\
AT          & 6.2MB   &  1.16$\times$   & \textbf{88.00$\times$} &    1.32$\times^*$    &     \textbf{29.29$\times$}  \\
\hline 
\end{tabular}
\end{table}

\begin{figure}
    \centering
    \includegraphics[width=\linewidth]{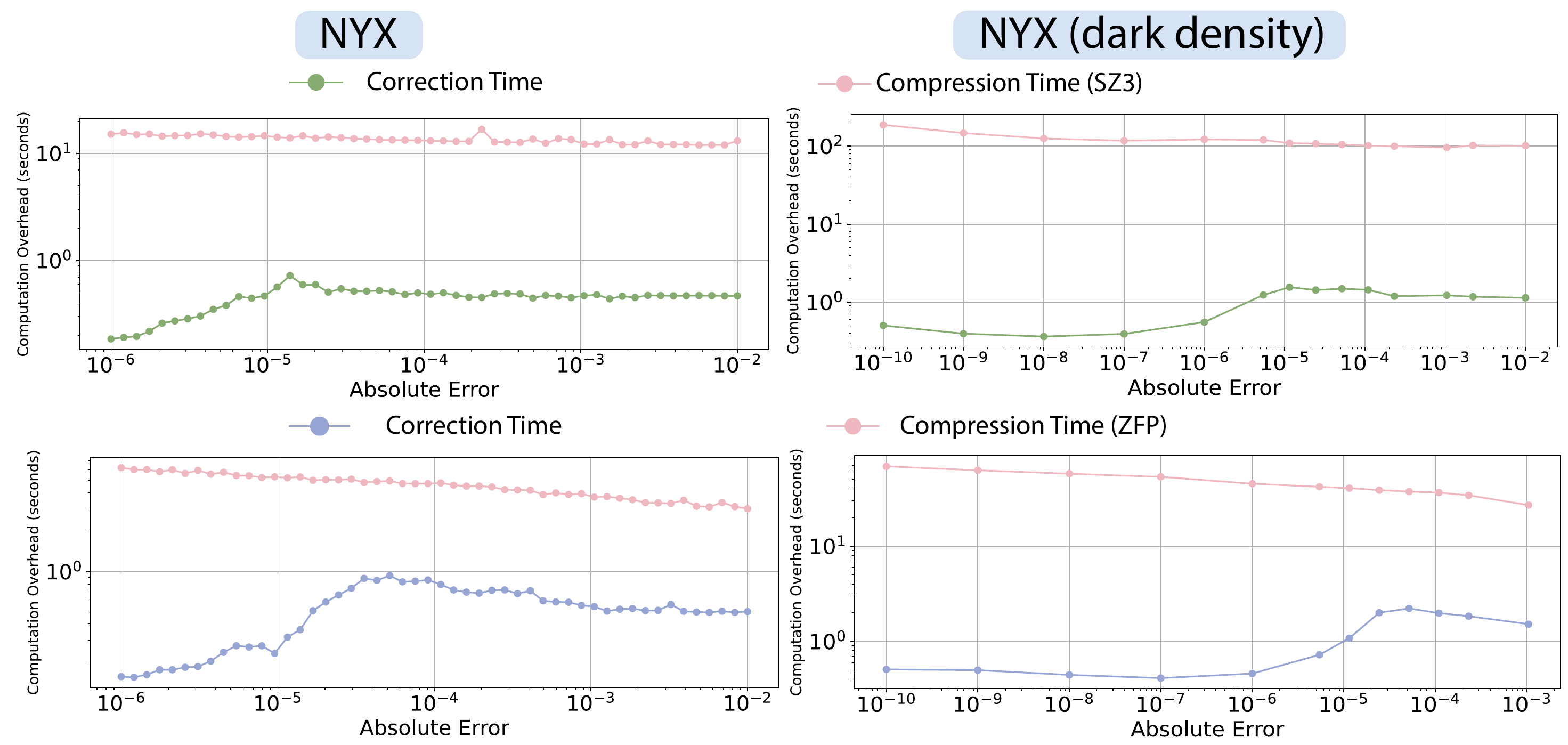}
    \caption{Computation overhead versus absolute error on the various datasets. Top: SZ3; bottom: ZFP. Pink lines represent the compression time of the base compressors.}

    \label{fig:overhead}
    \vspace{-0.1in}
\end{figure}

\noindent\textbf{Computation and Storage Overhead.}
Although our method introduces additional storage and computation overhead, the results in Figure~\ref{fig:overhead}, measured on a single node, show that we consistently achieve compression ratios up to $88\times$ under PLMSS preservation. 
In terms of computation overhead, the correction time of our method would not exceed the compression time, indicating that our method would not become a bottleneck in the compression pipeline.

For storage overhead, the additional cost comes from storing the edits. Nevertheless, even with storage cost, the achieved compression ratios remain substantially higher than those of the base compressors. For instance, on the Adenine Thymine (AT) dataset, our method reaches a CR of $88\times$ with SZ3, whereas the base compressor can only preserve PLMSS in a lossless mode, as shown in Table~\ref{tab:plmss_cr}. 

\begin{figure}[htb]
    \centering
    \includegraphics[width=\linewidth]{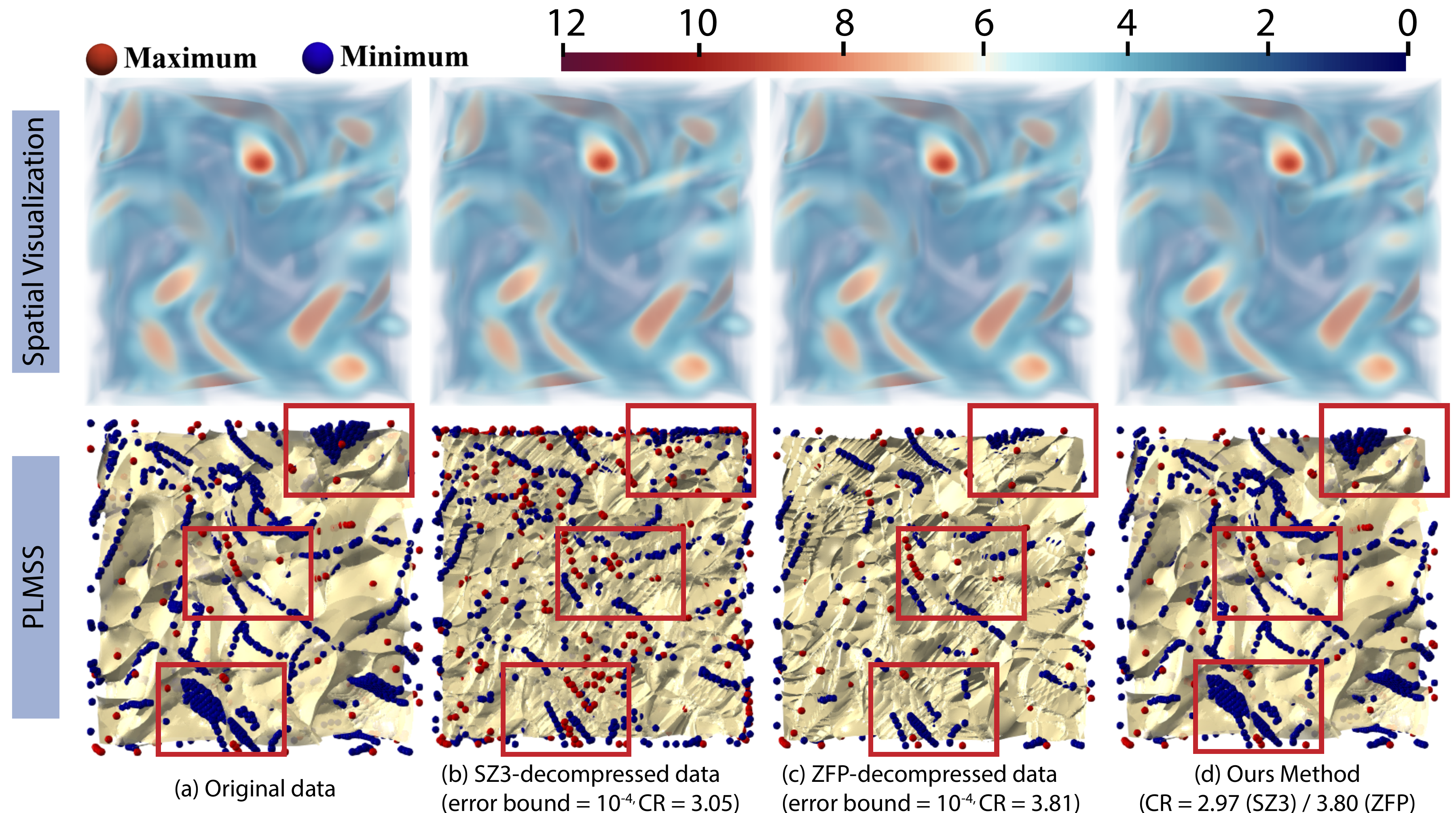}
    \caption{PLMSS of (a) Original data. (b,c) SZ3 and ZFP decompressed data under an error bound $10^{-4}$ with introduced distortions in the PLMSS (red boxes). (d) Our method on SZ3 and ZFP without PLMSS distortion.}
    \label{fig:vortex}
    \vspace{-0.2in}
\end{figure}

\noindent\textbf{Preservation of PLMSS.}
We use the vortex dataset to evaluate PLMSS preservation under a relative error bound of $10^{-4}$.
While the decompressed volumes from SZ3 and ZFP appear visually similar to the original, both introduce PLMSS errors: ZFP produces 3,082 wrongly labeled points, 8 false minima, 0 false maxima, and SZ3 produces 176,162 wrongly labeled points, 56 false maxima, and 153 false minima.
In contrast, our method achieves 100\% PLMSS preservation for both SZ3 and ZFP, correcting all PLMSS errors (Figure~\ref{fig:vortex}).

\section{Conclusion and Future Works }
In this paper, we address the primary scalability bottleneck of MSz by removing the computationally expensive R-loops that rely on repeated integral path computations. We propose a distributed-memory algorithm for PLMSS under error-bounded lossy compression. By reformulating the preservation condition to enable fully local corrections, our method achieves up to $14\times$ speedup over MSz on single-GPU execution. In addition, we design a relaxed interprocess communication strategy that allows each data block to reach local convergence before communicating with neighboring ranks, substantially reducing communication frequency and cost.
Leveraging CUDA-aware MPI, our distributed implementation efficiently scales to 128 GPUs with weak-scaling efficiency above 90\%, demonstrating that our method effectively addresses both computational and communication bottlenecks of MSz while maintaining 100\% PLMSS preservation.

Our work has several limitations. 
First, our method will introduce additional edits than MSz because it strictly preserves the local ordering of every data point. An adaptive policy that selectively relaxes ordering preservation further reduces edit counts and storage overhead.
Second, although we currently relaxed the ghost layer synchronization by allowing each block to reach local convergence before ghost exchanges, the boundary consistency may be delayed in certain cases. Future work could explore adaptive communication schemes to better balance accuracy and efficiency.
Third, by allowing local convergence before synchronization, different blocks may converge at different times because the distortion in different blocks might be different; early-converged blocks will remain idle while neighboring blocks continue local corrections. Incorporating asynchronous data communication could mitigate this issue and further improve scalability.

\section*{Acknowledgments}
This research is supported by the U.S. Department of Energy, Office of Science, Advanced Scientific Computing Research, under contract \texttt{DE-AC02-06CH11357}, and by the National Science Foundation under awards OAC-2313122, OIA-2327266, and OAC-2313124. This research used resources of the National Energy Research Scientific Computing Center (NERSC), a Department of Energy Office of Science User Facility.

\bibliographystyle{IEEEtran}
\bibliography{Refs}

\end{document}